\documentclass[runningheads]{llncs}
\usepackage{graphicx}
\usepackage{todonotes}
\usepackage{subfigure}
\usepackage[misc]{ifsym}
\usepackage{booktabs}
\usepackage{bm}
\usepackage{url}

\begin{document}
\title{Deep Learning Framework for Spleen Volume Estimation from 2D Cross-sectional Views }
\author{Zhen Yuan\inst{1}$^{(\textrm{\Letter})}$ \and
Esther Puyol-Antón\inst{1} \and
Haran Jogeesvaran\inst{2} \and
Baba Inusa\inst{2} \and
Andrew P. King\inst{1}}
%
\authorrunning{Zhen Yuan et al.}
%
\institute{School of Biomedical Engineering and Imaging Sciences, King's College London, London, UK 
\\
\email{zhen.1.yuan@kcl.ac.uk} \and
Evelina Children’s Hospital, Guy’s and St Thomas’ NHS Foundation Trust, London, UK}
\titlerunning{Deep Learning for Automatic Spleen Length Measurement}%
\maketitle              
\begin{abstract}
Abnormal spleen enlargement (splenomegaly) is regarded as a clinical indicator for a range of conditions, including liver disease, cancer and blood diseases. While spleen length measured from ultrasound images is a commonly used surrogate for spleen size, spleen volume remains the gold standard metric for assessing splenomegaly and the severity of related clinical conditions. Computed tomography is the main imaging modality for measuring spleen volume, but it is less accessible in areas where there is a high prevalence of splenomegaly (e.g., the Global South). Our objective was to enable automated spleen volume measurement from 2D cross-sectional segmentations, which can be obtained from ultrasound imaging. In this study, we describe a variational autoencoder-based framework to measure spleen volume from single- or dual-view 2D spleen segmentations. We propose and evaluate three volume estimation methods within this framework. We also demonstrate how 95\% confidence intervals of volume estimates can be produced to make our method more clinically useful. Our best model achieved mean relative volume accuracies of 86.62\% and 92.58\% for single- and dual-view segmentations, respectively, surpassing the performance of the clinical standard approach of linear regression using manual measurements and a comparative deep learning-based 2D-3D reconstruction-based approach. The proposed spleen volume estimation framework can be integrated into standard clinical workflows which currently use 2D ultrasound images to measure spleen length. To the best of our knowledge, this is the first work to achieve direct 3D spleen volume estimation from 2D spleen segmentations.

\end{abstract}
\section{Introduction}
\label{section1}
Splenomegaly refers to an abnormally enlarged spleen and is a clinical indicator for various conditions, such as liver disease, cancer, infections, and haematological disease \cite{mccormick2000splenomegaly,woodruff1973mechanisms,klein1987splenomegaly}. Spleen size is an essential biomarker for detecting splenomegaly and assessing the severity of associated clinical conditions. In the clinic, 2D ultrasound imaging is routinely used to identify splenomegaly by measuring the spleen length between the inferior and superior points in a longitudinal (approximately coronal) view. Although spleen length correlates well with spleen volume \cite{lamb2002spleen}, it is still considered a surrogate measure and spleen volume remains the gold standard for spleen size. In addition, apart from determining splenomegaly, spleen volume has more important diagnostic value compared to spleen length and can be regarded as a preferred indicator in determining the severity of diseases and informing treatment planning \cite{paley2002imaging,ref6}. Whilst spleen volume, in principle, can be estimated by performing linear regression on measurements from multiple ultrasound planes \cite{badran2015ultrasonographic,yetter2003estimating,ishibashi1991sonographic}, this approach is dependent on the observer’s experience and furthermore suffers from inter- and intra-observer variability. 3D modalities such as Computed Tomography (CT) provide the most robust estimates of spleen volume. Volume measurements from CT can be achieved either by linear regression based on manual measurements of length, width and thickness obtained from multiple 2D planes \cite{caglar2014determination} or by counting the number of voxels within a manually delineated spleen volume. The first approach lacks accuracy and the second is time-consuming. Deep learning-based frameworks have also been proposed for automatic spleen segmentation and volume estimation from CT images. For example, \cite{moon2019acceleration,ref12} both presented fully automatic pipelines to segment the spleen from CT images and then obtain the volume. However, despite its accuracy in spleen volume assessment, CT imaging involves ionising radiation, is expensive and less accessible in some parts of the world, especially in the Global South where there is a high prevalence of blood disease-related splenomegaly. As a result, CT imaging is not widely used in cases of suspected splenomegaly, and ultrasound remains the preferred modality, making the clinical value of spleen volume as a biomarker still under-investigated. Developing an accurate and robust framework to measure spleen volume that can potentially be integrated into routine ultrasound examination workflows is therefore important for improving the diagnosis and management of associated clinical conditions. 

In this study, we propose a novel deep learning-based framework for spleen volume estimation from 2D cross-sectional segmentations. Our overall objective is to develop a pipeline that can estimate spleen volume directly from 2D ultrasound images that are routinely acquired at standard clinical views. To address this objective, we take a two-stage approach: first, automatically segmenting the spleen from the 2D ultrasound images, and second estimating volume from the segmentations. We have previously published a robust and accurate method for automated spleen segmentation from 2D ultrasound \cite{yuan2022deep} and in this paper we address the second part of the problem: estimating volume from the 2D segmentations. The main contributions of our work can be summarised as: 
\begin{enumerate}
    \item We propose a novel variational autoencoder (VAE) based framework that can automatically estimate spleen volume from 2D spleen segmentations. 
    \item We propose and evaluate three different volume estimation methods within the framework.  
    \item As well as making an estimate of volume, our framework also allows estimation of 95\% confidence intervals to support decision-making in real clinical scenarios. 
    \item The performance of our framework surpasses the current clinical standard approach and a comparative deep learning-based 2D-3D reconstruction-based method in estimating spleen volume.
    \item To the best of our knowledge, this is the first work to achieve direct 3D spleen volume estimation from 2D spleen segmentations.
\end{enumerate}

\section{Related Work}
\label{section2}
\textbf{3D Segmentation and Direct Volume Estimation.} Deep learning-based methods have been proposed to automate volume measurements through automatic segmentation of 3D imaging in organs such as the heart \cite{liao2017estimation}, liver \cite{wang2019automated} and kidney \cite{sharma2017automatic}. In addition, segmentation-free volume estimation models have been developed for direct regression of ventricular volume \cite{zhen2015direct} and kidney volume \cite{hussain2021cascaded}, again from 3D imaging. In the spleen, 3D convolutional neural networks (CNNs) were proposed by \cite{moon2019acceleration,ref12} to segment the spleen from 3D CT images and subsequently estimate its volume, whilst \cite{huo2018splenomegaly} focused on spleen segmentation from splenomegaly magnetic resonance images (MRI). The works mentioned above all require 3D modalities for volume estimation. 

\textbf{2D-to-3D Reconstruction.} In computer vision methods have been proposed to estimate 3D shape from 2D images. These include recent works like Pix2Vox++ \cite{xie2020pix2vox++}, which learnt to reconstruct 3D shape from 2D images and 3DAttriFlow \cite{wen20223d}, which utilising a point cloud and 2D images to estimate 3D shape. Prior to the emergence of deep learning, 2D-to-3D reconstruction was investigated in the medical imaging domain, for example in \cite{baka20112d}, which proposed a method based on statistical shape models to reconstruct 3D shape from 2D X-ray images. Pix2Vox++ has also been applied in medical imaging, specifically to 3D cardiac reconstruction in \cite{stojanovski2022efficient}. However, reconstruction-based methods often require manual clinical landmarks to select salient 2D slices, and good reconstruction quality is normally only possible with multiple views (more than two) acquired from different angles. For spleen volume estimation this would present a problem. The relatively low accuracy of 3D reconstruction from a more limited number of views could lead to inaccurate volume estimation, whilst obtaining more than two ultrasound views could be challenging due to the limited acoustic windows. 

\textbf{VAE-based Approaches.} VAEs are able to learn a low-dimensional representation (or latent space) from high-dimensional data such as images or segmentations, and provide benefits in terms of interpretability by allowing reconstruction to the original (image) space from their latent space. VAEs have been previously combined with a secondary task in a similar way to our framework. For example, \cite{biffi2018learning} proposed a VAE based framework which also performed disease diagnosis from cine cardiac MRI derived segmentations. The use of a VAE enabled interpretability by visualising the learned features for healthy and hypertrophic cardiomyopathy patients. Similarly, in \cite{puyol2020interpretable}, a residual block-based VAE was trained using cine cardiac MRI segmentations to perform classification for predicting treatment response as well as explanatory concepts. In \cite{puyol2020assessing}, a VAE based on cardiac functional biomarkers was combined with a regression task in the latent space to learn the relationship between cardiac function and systolic blood pressure. \cite{cetin2023attri} proposed the Attri-VAE model to enable disentangled interpretation based on clinical attributes between images of healthy and myocardial infarction patients. In \cite{zhao2019variational}, a conditional generative VAE-based model was used to enable direct regression of brain age from 3D MRI images. \cite{joo2023variational} proposed a framework that utilised VAEs to learn features from various teeth images, and subsequently applied linear regression to estimate the corresponding subjects’ age. To the best of our knowledge, VAEs have not yet been investigated for estimation of volume from 2D views. 

\section{Materials}
\label{section3}
\subsection{Data Description}
\label{section3.1}
Our dataset is comprised of 149 manual segmentations of computed tomography (CT) volumes from two publicly available collections, with 36 exhibiting the characteristic of splenomegaly (characterised by a volume greater than 314.5mL) and the remaining 113 having a volume that falls within a normal range \cite{prassopoulos1997determination}. 60 spleen segmentations were accessed from the Medical Segmentation Decathlon (MSD) challenge, acquired at the Memorial Sloan Kettering Cancer Centre, New York, US \cite{antonelli2022medical}. This dataset consists of 40 3D CT volumes with associated manual segmentations (41 cases in total, but one image lacks part of the spleen and was therefore excluded from our experiments) and 20 CT volumes without manual segmentations. An experienced radiologist performed additional manual segmentations of these 20 CT volumes, resulting in a final dataset of 60 manual segmentations of CT volumes from MSD. The second data source is from Gibson et al. \cite{gibson2018automatic}, which consists of 90 CT volumes with manual segmentations of multiple abdominal organs including the spleen. This dataset is a combination of 43 subjects from the Cancer Imaging Archive Pancreas-CT dataset \cite{roth2015deeporgan,clark2013cancer} and 47 subjects from the ‘Beyond the Cranial Vault’ (BTCV) segmentation challenge \cite{MICCAI}. The Pancreas-CT dataset was acquired from either healthy kidney donors prior to nephrectomy or patients who neither had major abdominal pathologies nor pancreatic cancer lesions. The BTCV scans were randomly selected from a combination of an ongoing cancer chemotherapy trial and a retrospective ventral hernia study. One BTCV segmentation was excluded due to an excessively large spleen volume (3083mL), which was considered an outlier even for a case of splenomegaly, as the average spleen volume for a splenomegaly case has been determined to be 1004.75 ± 644.27mL \cite{linguraru2013assessing}. Thus, in total, 89 segmentations from Gibson et al. were utilised in our study, and all manual spleen segmentations were obtained directly from the publishers. Note that for all datasets only the corresponding spleen segmentations of CT volumes were utilised in our work, not the original CT volumes themselves. 

\subsection{Data Pre-processing }
\label{section3.2}
A series of pre-processing operations were conducted to ensure uniformity among the 149 segmentations obtained from the different sources. First, labels for abdominal organs other than the spleen were removed from the Gibson et al. dataset, followed by lateral flipping to align with the orientation of segmentations from the MSD challenge dataset. We resampled all volumes to a uniform voxel size of $1\times 1\times 1 mm^{3}$ using nearest neighbour interpolation. Segmentations were then cropped based on calculated bounding boxes and centroids. Specifically, we translated the segmentations so that their centroids aligned, followed by padding to a size of $164\times 186\times 176$ to maintain consistency according to the largest bounding box across the dataset. Finally, we processed the segmentations using a mode filter with size of 7 along the coronal plane to smoothen the spleen boundary.  

\begin{figure}[htb] 
\centering 
\includegraphics[width=0.7\textwidth]{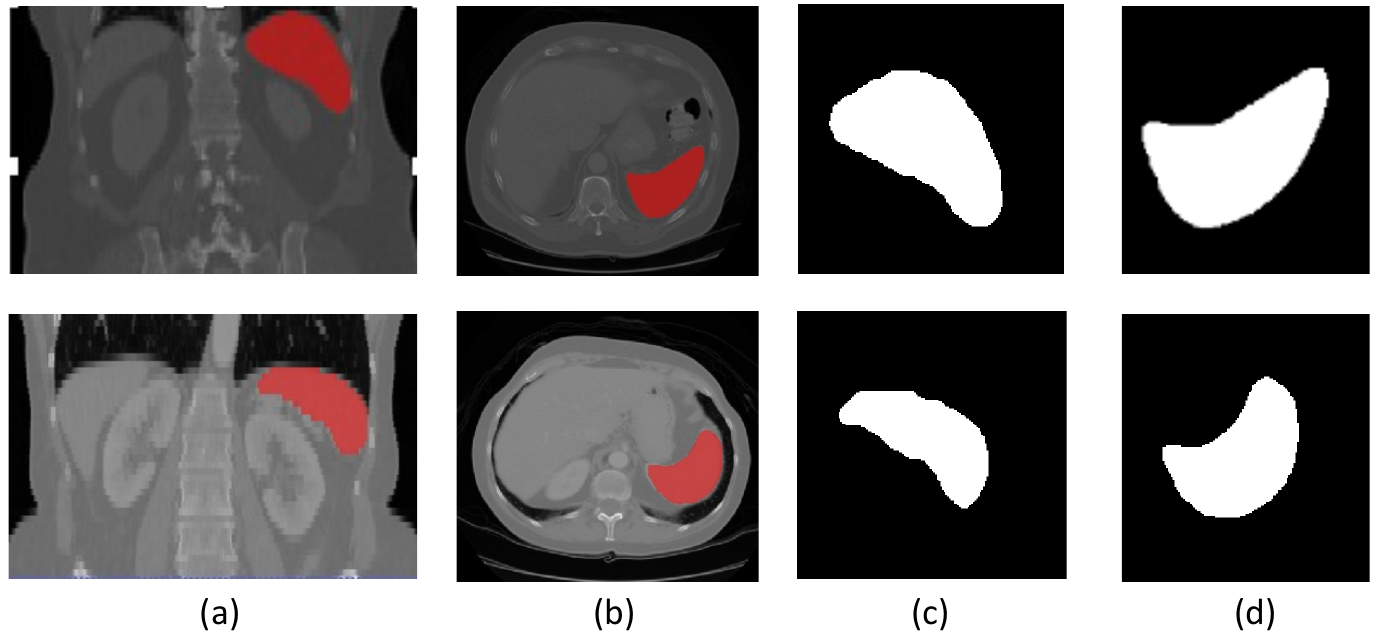} 
\caption{Two example CT volumes (top and bottom rows) with associated ground truth spleen segmentations and the selected coronal and transverse 2D segmentation slices. (a). Coronal view of CT volume with spleen segmentation in red. (b). Transverse view of CT volume with spleen segmentation in red. (c). Selected coronal 2D segmentation slice with largest cross-sectional area. (d). Selected transverse 2D segmentation slice with largest cross-sectional area.} 
\label{fig1} 
\end{figure}

After combining the segmentations from the two sources, we extracted 2D segmentations from the 3D segmentations. For each 3D segmentation we selected the central coronal and transverse slices with the largest cross-sectional areas. This resulted in two 2D segmentation slices for each CT volume. To ensure consistency in size, all 2D slices were then resampled to a size of $224\times 224$. As a result, we obtained 149 pairs of coronal and transverse 2D segmentations, derived from 149 3D spleen segmentations. All ground truth volumes for the spleen were calculated from the 3D segmentation as the product of the number of voxels within the spleen and the voxel size. Fig. \ref{fig1} presents example 3D CT volumes with their associated ground truth segmentations and the corresponding central coronal and transverse segmentation slices.

\section{Methods}
\label{section4}
In this section, we present our VAE-based approach for automated estimation of volume from 2D single or dual view spleen segmentations.   

We first introduce the architecture of the proposed framework in section \ref{section4.1} and then describe three different methods for automatic volume estimation in section \ref{section4.2}. Finally, we provide details of an adapted framework that harnesses the generative nature of the VAE to provide estimated volumes with associated confidence intervals, which is aimed at improving the model’s utility for clinical end-users (see section \ref{section4.3}).  

\subsection{Network Structure}
\label{section4.1}
\begin{figure}[htb] 
\centering 
\includegraphics[width=1\textwidth]{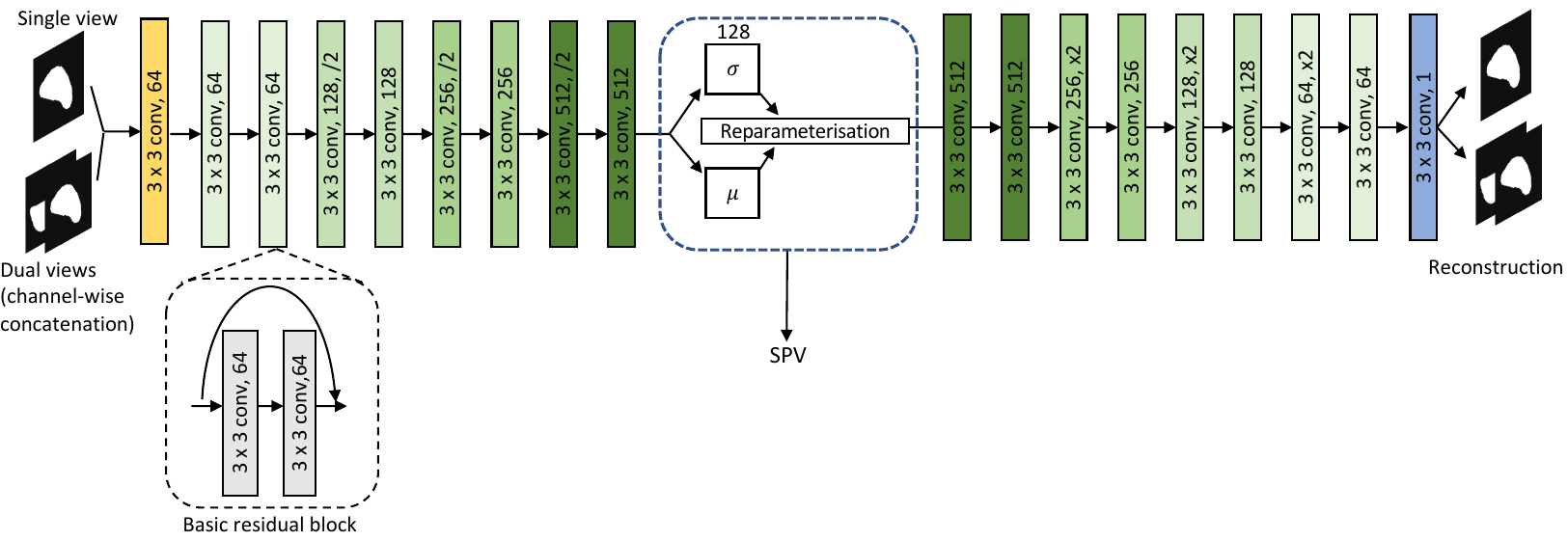} 
\caption{An illustration of the proposed VAE-based framework. The latent space distribution is parameterised by mean $\mu$ and standard deviation $\sigma$, which is shown in the blue dotted frame. The spleen volume is estimated from this distribution. The basic residual block is shown in the black dotted frame. All our VAE-based models were trained with either a coronal slice (single view) or a channel-wise concatenation of a coronal slice and a transverse slice (dual views). SPV: spleen volume.} 
\label{fig2} 
\end{figure}

Our framework was based upon a residual block-based VAE. The framework consisted of an encoder and a decoder, where the encoder compressed the input into a latent space distribution and the decoder restores the data from the low-dimensional representation in the latent space back to the input space. The architecture of the proposed network is depicted in Fig. \ref{fig2}.

We utilised the residual block due to its powerful capabilities in the medical image processing domain. The encoder and decoder both consisted of 8 basic residual blocks, each of which comprised two cascades of convolutional layer, batch normalisation and ReLU activation function (see the detail of basic residual blocks in Fig. \ref{fig2}). Input and output layers were deployed to process the input segmentations and output features. The mean $\mu$ and standard deviation $sigma$ parameterise the latent space distribution, each with a dimension of 128. During training, the encoder learnt to map inputs to the latent space by sampling based on the $\mu$ and the $\sigma$, and the decoder restored the segmentation using the reparametrised representation $z=\mu+  \zeta \odot \sigma$, where $\zeta \sim N\left ( 0,I \right ) $ is a noise vector and $\odot$ represents element-wise multiplication. 

The basic loss function we used to optimise the VAE comprised binary cross entropy calculated between original label A and reconstructed label B and Kullback-Leibler divergence loss to regularise the latent distribution weighted by $w_{1} $.

\begin{equation}
loss= BCELoss(A,B)+  w_{1} KLDloss \label{eq1}
\end{equation}

Note that all models were trained in two settings – with only a coronal slice (single view) and also with a channel-wise concatenation of coronal and transverse slices (dual views). See section \ref{section5} for details. 

\subsection{Volume Estimation}
\label{section4.2}
We propose and investigate three different methods to estimate spleen volume from the latent space distribution.  

\textbf{Nearest neighbour searching in the latent space (NN).}  In this method, we first trained the VAE to model the distribution of observed segmentations. Then, during inference, we mapped the input segmentation into the latent space and searched for the nearest neighbour from the training dataset. The corresponding ground truth volume of this nearest neighbour was used as the volume estimate for the test case. 

Formally, we denote by $L^{I} $ the slice or slices from the test segmentation, which can be a single coronal slice, or a channel-wise concatenation of a coronal slice and a transverse slice. During inference, we inputted $L^{I} $ into the VAE, resulting in the latent space representation mean $\mu ^{I} $, whose dimension is 128. We then inputted each sample in the training dataset $L^{T}=\left \{ L_{1}^{T},L_{2}^{T},L_{3}^{T},L_{4}^{T},L_{5}^{T},...,L_{N}^{T}  \right \}  $ into the trained VAE and get representations $\mu^{T}=\left \{ \mu_{1}^{T},\mu_{2}^{T},\mu_{3}^{T},\mu_{4}^{T},\mu_{5}^{T},...,\mu_{N}^{T}  \right \}  $. We calculated the Euclidean distance between each $\mu_{n}^{T} $ and $\mu^{I} $, and used the closest training sample’s corresponding spleen volume as the estimated volume for the test subject. 

\begin{figure}[!htb] 
\centering 
\includegraphics[width=1\textwidth]{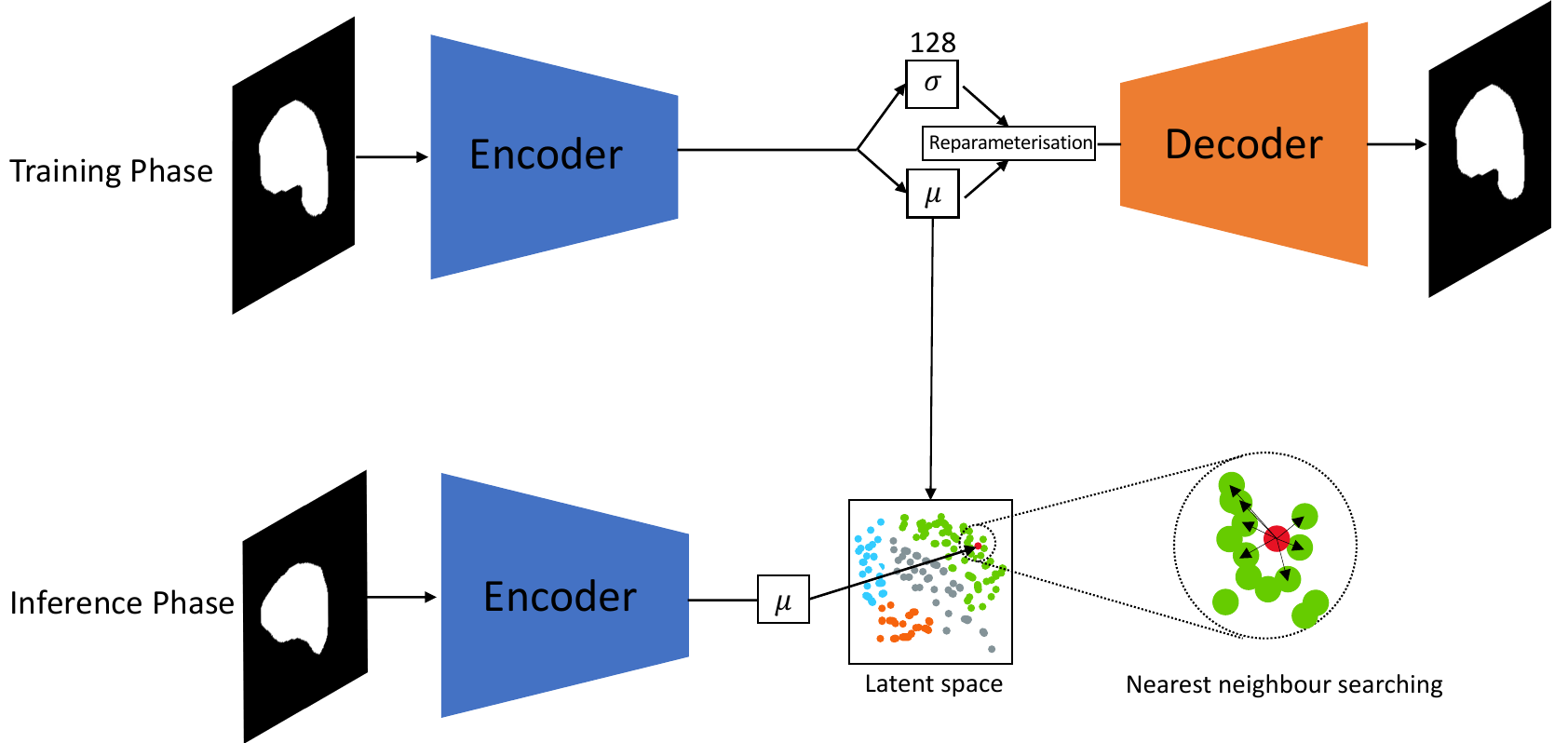} 
\caption{Illustration of nearest neighbour searching in the latent space (NN). In the training phase (see the upper part of the figure), the VAE learns to compress and reconstruct the 2D segmentation slices. In the inference phase (the lower part of the figure), the VAE maps the unseen segmentation slice into the latent space. We searched for the nearest neighbour in the latent space for each input by calculating the Euclidean distance. } 
\label{fig3} 
\end{figure}

In this method we used the basic loss function presented in Eq. \ref{eq1} to train the VAE. An illustration of this volume estimation approach is shown in Fig. \ref{fig3}. 

\clearpage

\begin{figure}[!htb] 
\centering 
\includegraphics[width=1\textwidth]{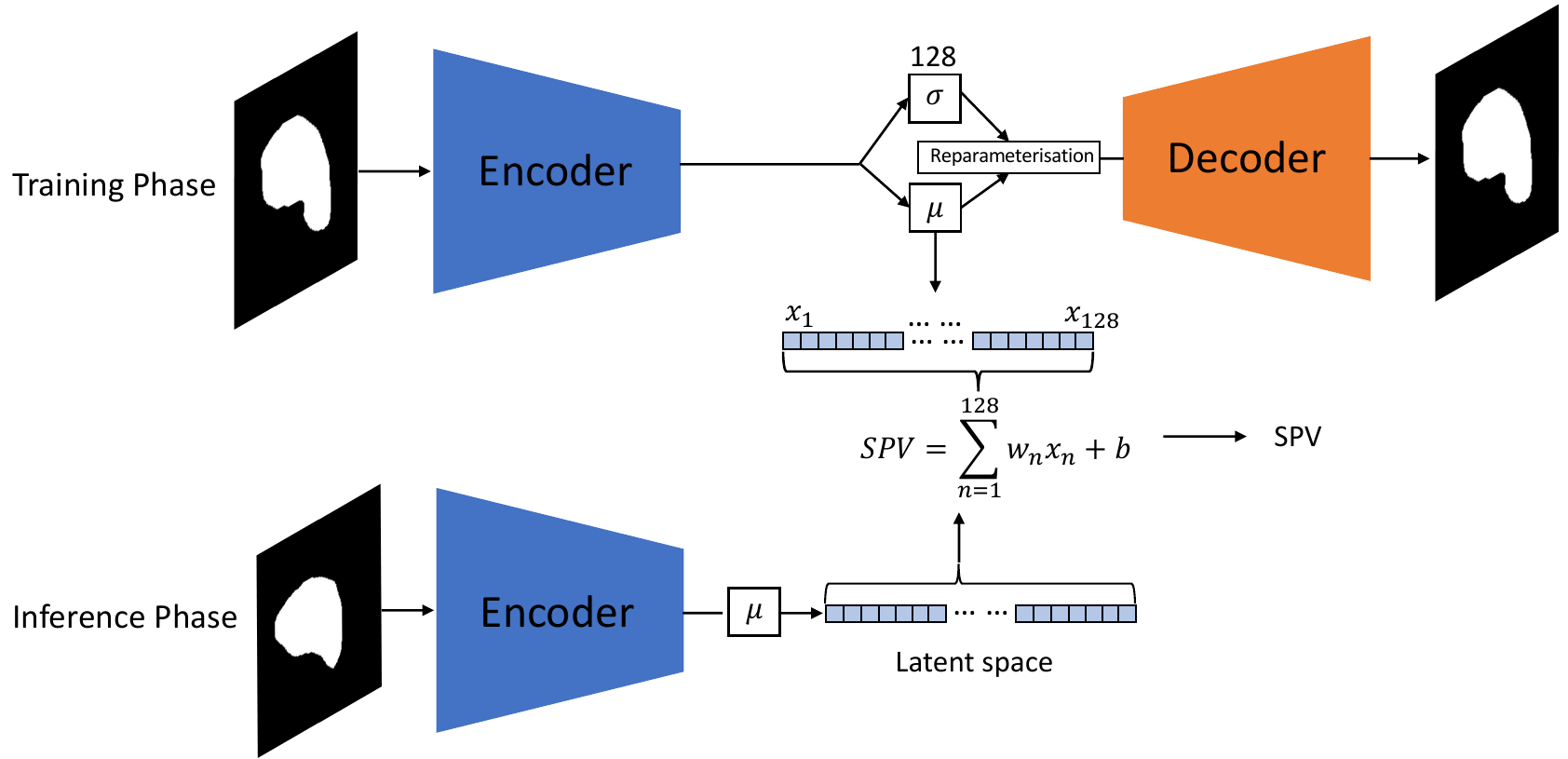} 
\caption{An illustration of linear regression of latent space coordinates to predict the volume (LR). We obtained the linear regression equation from the embeddings of training images in the latent space. In the inference stage, we predicted spleen volume using the test sample embedding and the obtained regression equation. } 
\label{fig4} 
\end{figure}

\textbf{Post linear regression of latent representations (PLR).}  In this method, after the VAE was trained using the basic loss function presented in Eq. \ref{eq1}, a simple linear regression was performed on the $\mu$ of the training data to predict the spleen volume. At test time, each slice was embedded into the latent space and its 
$\mu$ value used with the obtained regression equation to estimate the volume. Fig. \ref{fig4} illustrates the proposed PLR method. 

\begin{figure}[!htb] 
\centering 
\includegraphics[width=1\textwidth]{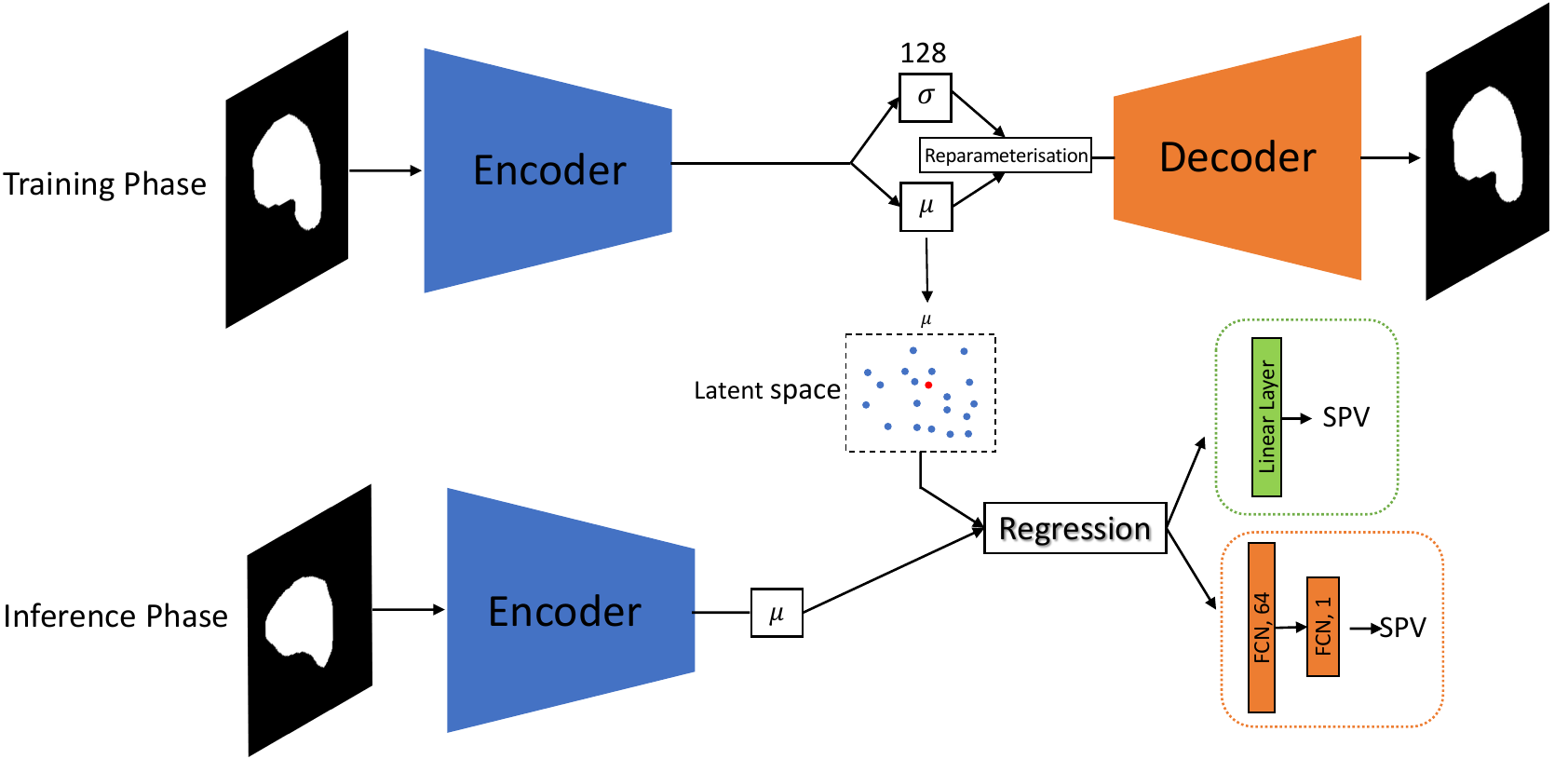} 
\caption{An illustration of RVAE models with RVAE-LR in green dotted frame and RVAE-FCNR in orange dotted frame. In RVAE-LR and RVAE-FCNR, during the training stage and inference stages, the mean $\mu$ embeddings were inputted into a linear layer and fully connected layers to make predictions for spleen volume (SPV). } 
\label{fig5} 
\end{figure}

\textbf{End-to-end regression VAE (RVAE). } This method was also based upon regression of volume from the VAE latent space, but here we incorporated additional layer(s) into the VAE to predict the spleen volume from the latent space in an end-to-end training manner. We implemented the RVAE method in two forms – using a linear regression layer only (RVAE-LR) and using fully connected layers with activation functions to model the potentially non-linear relationship between the latent embeddings and volume (RVAE-FCNR). In the proposed RVAE-LR model, a simple linear regression layer was incorporated to estimate the spleen volume, formulated by $vol^{'}=W^{T}  \mu +  b$, where $vol^{'}$ denotes estimated volume, and $W$ and $b$ are the regression coefficients of the obtained linear equation. In the RVAE-FCNR model, we added a fully connected layer with size 64 and ReLU activation function, followed by an output layer. For both RVAE-LR and RVAE-FCNR, during training, we included an extra regression term to the basic loss function in Eq. \ref{eq2} to encourage the network to learn to estimate the spleen volume. 

\begin{equation}
    loss=BCEloss(A,B)+w_{1} KLDloss+  w_{2} MSEloss(vol,vol^{'} ) \label{eq2}
\end{equation}

\noindent where A and B denote the input segmentation and reconstructed segmentation, $vol$ is the ground truth volume, $vol^{'}$ is the estimated volume and $w_{1}$ and $w_{2} $ are weights for the Kullback-Leibler divergence and mean squared error losses. An illustration of the RVAE methods is shown in Fig. \ref{fig5}. 

\subsection{Volume Estimation with Confidence Intervals}
\label{section4.3}
To further harness the generative nature of VAEs, we also integrated a method to compute 95\% confidence intervals in addition to a scalar estimated volume. Such information could be beneficial to clinicians in evaluating possible splenomegaly using ultrasound and allow greater flexibility in the use of our method as a decision support tool. We implemented this approach only for the RVAE-FCNR method due to the similar results of the RVAE-based methods (see sections \ref{section5} and \ref{section6}). In addition, the linear regression layer used in RVAE-LR was not constrained by activation functions, which can lead to negative volume predictions. In order to train the RVAE-FCNR network to make multiple predictions for a single input, we modified the input to the fully connected layers $\mu$ to the reparametrised representation $z=\mu+  \zeta \odot \sigma$ in the training stage, which is the same as the input provided to the decoder of the VAE. At inference time, we sampled in the latent space 100 times using $\zeta \sim N\left ( 0,I \right ) $. For each sampled reparametrised representation z, we inputted it into the fully connected layers to estimate the spleen volume using the method described above. The 95\% confidence interval was then determined by $\eta \pm 1.96\times \theta $, where $\eta$ and $\theta$ are the mean and standard deviation of the 100 volume estimates. We denote this method by RVAE-FCNR-CI. 

\section{Experimental Details}
\label{section5}
\subsection{Volume Estimation and Confidence Intervals}
\label{section5.1}
As explained in section \ref{section3}, we employed 149 segmentation volumes from two distinct sources to develop our proposed volume estimation methods. For each 3D segmentation, we generated two types of 2D segmentations – one with only a coronal slice (single view) and the other with the combination of a coronal slice and a transverse slice (dual views). For each proposed volume estimation method, we trained models under the single-view and dual-view scenarios, resulting in two sets of results for each of the proposed models.  

To evaluate the performance of our models, we split the dataset into 5 folds, each fold containing 30 segmentations. Two folds contained a duplicated case, but it was not included in the hold-out test fold. Each fold had 7-8 segmentations manifesting splenomegaly (to be more specific, 7 for four folds and 8 for one fold), i.e., its volume was greater than 314.5mL. One fold (30 segmentations including 7 splenomegaly cases) was regarded as the hold-out test set, and to analyse the proposed methods comprehensively, we conducted a 4-fold stratified cross-validation on the remaining folds, with 3 folds used for training and 1 fold used for each validation. The evaluation of each method was based on the average performance of the 4 models trained from this cross-validation on the hold-out test set.  

We employed different training strategies for different methods. Specifically, for NN and PLR, we trained the model for up to 500 epochs with $w_{1} $ set to 0.2. For the RVAE-based methods, i.e. RVAE-LR, RVAE-FCNR, and RVAE-FCNR-CI, we first trained models using only the VAE loss (i.e., binary cross entropy and the Kullback-Leibler divergence loss) with $w_{1}=0.2$ and $w_{2}=0$ for 150 epochs, and then trained them with both VAE loss and regression loss ($w_{1}=0.2$ and $w_{2}=0.2$) for up to 650 epochs. For each evaluated method and each cross-validation fold, we used a grid search strategy for values from 0.1 to 0.5 in steps of 0.1 using its validation set to select the optimal values of the hyperparameters $w_{1} $ and $w_{2} $, considering the trade-off between volume estimation, disentanglement, and reconstruction quality. We also conducted hyperparameter optimisation for learning rate (values 0.0001, 0.001 and 0.01) and minibatch size (values 4 and 8). Based on this, we set learning rate = 0.001 and minibatch size = 8 for all methods. 

During the training phase of all methods, we performed data augmentation by applying small random rotations ranging from -15 to +15-degrees along all three axes to the 3D segmentations prior to the selection of the coronal and transverse slices. We saved the final trained models as those with the best volume estimation performance on the validation set. All models were trained on a NVIDIA GeForce GTX TITAN X 24GB using the Adam optimiser. Note that the ground truth volumes were scaled down by a factor of 10 for use in training/inference, and the estimated volumes were scaled back up by a factor of 10 after inference. 

\subsection{Comparative Evaluation}
\label{section5.2}
For comparative evaluation, the Pix2Vox++ 2D-to-3D reconstruction framework \cite{xie2020pix2vox++} and clinical approach of linear regression on manual measurements were implemented \cite{caglar2014determination,bezerra2005determination}. Note that, rather than estimating volume directly, the Pix2Vox++ method estimates a 3D segmentation, from which we computed the volume. Pix2Vox++ includes two versions: a version with lower computational complexity called Pix2Vox++/F and an advanced version with higher computational complexity called Pix2Vox++/A. Both frameworks have an encoder, decoder, and a merger, while the Pix2Vox++/A has an additional refiner. We evaluated both versions on our data. 

The same data stratification and cross-validation approaches described in section \ref{section5.1} were applied for the comparative approaches. The Pix2Vox++ models were trained to reconstruct 3D segmentations from both single-view data or dual-view data, and we obtained the spleen volume by the product of the number of the voxels within the spleen and the voxel size. Following the process in the original paper, we resized the 3D ground truth volume to $32\times32\times32 $ for use in training/inference. The estimated volume was obtained by rescaling the reconstructed 3D segmentation from $32\times32\times32 $ back to $164\times186\times176 $. We trained both models with a learning rate of 0.001 (chosen from 0.0001, 0.001 and 0.01) and a minibatch size of 8 (chosen between 4 and 8) after hyperparameter optimisation.  

\begin{figure}[!htb] 
\centering 
\includegraphics[width=0.6\textwidth]{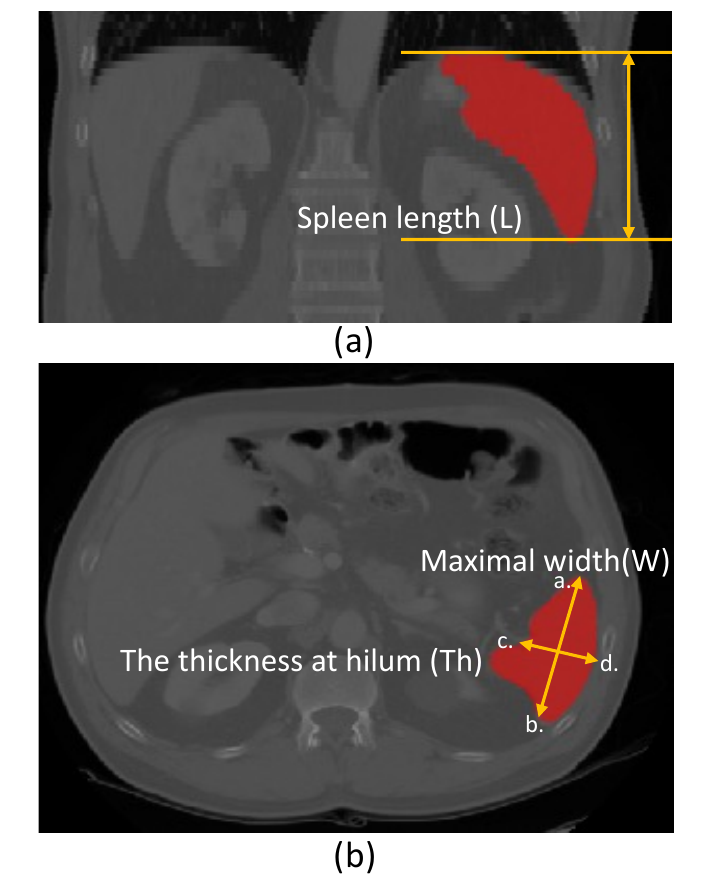} 
\caption{ Illustration of manual measurements from CT images for spleen volume estimation. (a). Coronal view. The spleen length (L) is obtained by multiplying the spacing of the coronal plane by the number of contiguous slices that contain the spleen along the transverse axis. (b). Transverse view. The maximal width (W) is obtained by finding the largest diameter on any transverse slice (see distance between ab). The thickness at hilum (Th) is determined by the thickness of the spleen at hilum and perpendicular to the spleen width (see distance between cd).  } 
\label{fig6} 
\end{figure}

In addition to Pix2Vox++, we also compared our proposed models with the approach most commonly used in current clinical workflows. This approach is based on linear regression of volume from manual 2D measurements of spleen size. The first variant used a single measurement of maximal spleen length based on the equation proposed in \cite{bezerra2005determination}, and the second variant used three orthogonal measurements (spleen length, maximal width and thickness at the spleen hilum) based on the equation proposed in \cite{caglar2014determination}. To obtain the spleen length, a human expert manually identified the most superior and inferior transverse slices that contained the spleen and multiplied the number of the slices by the thickness of the transverse slices. The human expert also measured the maximal width and the thickness at hilum, where the maximal width of the spleen was defined as the largest diameter on any transverse slice and the thickness at hilum was determined by the thickness of the spleen at hilum and perpendicular to the spleen width. An example of the manual measurement process is illustrated in Fig. \ref{fig6}. 

\section{Results}
\label{section6}
\subsection{Evaluation Metrics}
\label{section6.1}
To evaluate the performance of the proposed models for spleen volume estimation, we calculated the mean relative volume accuracy (MRVA) and Pearson’s correlation coefficient (R) as follows. 

\begin{equation}
    RVA= \frac{1}{N} \sum_{n=1}^{N} \left ( 1- \frac{\left | vol_{n}^{'}-vol_{n}   \right | }{vol_{n} }\times 100\%  \right ) \label{eq3}
\end{equation}

\begin{equation}
    \rho _{V^{'},V } =\frac{cov\left ( vol^{'},vol  \right ) }{\sigma _{vol^{'},vol } }  \label{eq4}
\end{equation}

\noindent where $V=\left \{ vol_{1}, vol_{2},vol_{3},...,vol_{N}, \right \} $ denotes the ground truth volumes and $V^{'} =\left \{ vol_{1}^{'}, vol_{2}^{'},vol_{3}^{'},...,vol_{N}^{'} \right \} $ denotes estimated volumes.  

We also proposed a confidence interval accuracy (CIA) metric to evaluate how well the confidence limits captured the uncertainty inherent in the volume estimates. CIA is defined as the percentage of predictions for which the ground truth volumes are within the estimated 95\% confidence limits. 

To assess the capability of our model to correctly distinguish splenomegaly cases, we also used three metrics: sensitivity (SEN), specificity (SPE) and accuracy (ACC): 

\begin{equation}
    SEN=\frac{TP}{TP+FN}  \label{eq5}
\end{equation}

\begin{equation}
    SPE=\frac{TN}{TN+FP}  \label{eq6}
\end{equation}

\begin{equation}
    ACC=\frac{TP+TN}{TP+TN+FP+FN}  \label{eq7}
\end{equation}

\noindent where TP refers to the number of splenomegaly cases with an estimated volume that is greater than 314.5mL, TN refers to the number of normal cases with an estimated volume that is smaller than 314.5mL, FN is the number of splenomegaly cases with an estimated volume that is smaller than 314.5mL and FP is the number of normal cases with an estimated volume that is greater than 314.5mL. 

\subsection{Experimental Results}
\label{section6.2}
For the volume estimation experiments, the results of training with single-view data (coronal slice only) and dual-view data (coronal and transverse slices) are shown in Table \ref{tab1}. “HE” in Table \ref{tab1} denotes the clinical approach of linear regression for spleen volume from manual measurements, using either length only (for the single-view data) or the 3 measurements (for the dual-view data), as performed by a human expert (HE).

\begin{table}[!htb]
\caption{Comparison of results between NN, LR, RVAE-LR, RVAE-FCNR and Pix2Vox++ trained on coronal single-view data and dual view (coronal and transverse) data. HE denotes spleen volume estimated by human expert using manual linear regression. Mean relative volume accuracy (MRVA) and standard deviation (STD). R: Pearson’s correlation coefficient. SEN: sensitivity. SPE: specificity. ACC: accuracy. Best results are in bold. }\label{tab1}
\centering
\begin{tabular}{@{}cccccc@{}}
\toprule
\multicolumn{3}{c}{\textbf{Volume}}                    & \multicolumn{3}{c}{\textbf{Splenomegaly}}              \\ \midrule
\textbf{}   & \bm{$MRVA\pm STD$} & \textbf{R}      & \textbf{SEN}     & \textbf{SPE}     & \textbf{ACC}     \\
\multicolumn{6}{c}{\textbf{Single view}}                                                                        \\
NN          & $73.58\%\pm 24.61\%$   & 0.8046          & 71.43\%          & 89.13\%          & 85.00\%          \\
PLR         & $53.76\%\pm 65.60\%$   & 0.4742          & 60.71\%          & 83.70\%          & 78.33\%          \\
RVAE-LR     & $86.58\%\pm 10.77\%$   & \textbf{0.9409} & \textbf{85.71\%} & \textbf{98.91\%} & \textbf{95.83\%} \\
RVAE-FCNR   & \bm{$86.62$}$\%\pm 11.37\%$   & 0.9406          & \textbf{85.71\%} & 94.57\%          & 92.50\%          \\
Pix2Vox++/F & $82.51\%\pm 12.33\%$   & 0.9244          & 67.86\%          & 97.83\%          & 90.83\%          \\
Pix2Vox++/A & $82.83\%\pm$ \bm{$ 10.74\%$}   & 0.9258          & 67.86\%          & \textbf{98.91\%} & 91.67\% \\
HE          & $68.54\%\pm 23.62\%$   & 0.8423          & -                & \textbf{-}       & -                \\
\multicolumn{6}{c}{\textbf{Dual views}}                                                                         \\
NN          & $85.82\%\pm 12.60\%$   & 0.9317          & 92.86\%          & 100\%            & 98.33\%          \\
PLR         & $84.40\%\pm 15.70\%$   & 0.8773          & 71.43\%          & 96.74\%          & 90.83\%          \\
RVAE-LR     & $92.53\%\pm 6.14\%$    & \textbf{0.9889} & \textbf{100\%}   & \textbf{100\%}   & \textbf{100\%}   \\
RVAE-FCNR   & \bm{$92.58\%$} $\pm$ \bm{$6.07\%$}    & 0.9766          & 92.86\%          & 98.91\%          & 97.50\%          \\
Pix2Vox++/F & $85.28\%\pm 9.52\%$    & 0.9152          & 67.86\%          & 96.74\%          & 90.00\%          \\
Pix2Vox++/A & $86.27\%\pm 10.30\%$   & 0.9293          & 82.14\%          & 93.48\%          & 90.83\%          \\
HE          & $81.60\%\pm 14.50\%$   & 0.9643          & -                & -                & -                \\ \bottomrule
\end{tabular}
\end{table}

It can be seen from Table \ref{tab1} that our RVAE-based models (RVAE-LR and RVAE-FCNR) surpassed human expert performance (MRVA 68.54\% and 81.60\% for single/dual view experiments) and the comparative 2D-to-3D reconstruction-based framework Pix2Vox++ (best MRVA 82.83\% and 86.27\% for single/dual view experiments) in the task of volume estimation from 2D slices. The RVAE-based methods also outperformed the NN and PLR methods. The RVAE-FCNR method achieved slightly better volume estimation accuracy than RVAE-LR, with a MRVA of 86.62\% compared to 86.58\% when trained with single-view data, and a MRVA of 92.58\% compared to 92.53\% when trained with dual-view data. However, RVAE-LR had a slightly better performance in splenomegaly prediction compared to RVAE-FCNR. Pix2Vox++/F and Pix2Vox++/A had similar results for both experiments (82.51\% and 82.83\ MRVA trained with single-view data, 85.28\% and 86.27\% MRVA trained with dual-view data). 

The performance of all approaches was improved when making use of dual-view data. The PLR model exhibited the largest rise from 53.76\% to 84.40\% in MRVA. In contrast, the comparative framework Pix2Vox++ did not show noticeable improvement. There was a substantial improvement in splenomegaly prediction for all our proposed models when making use of dual-view data.  

The results for RVAE-FCNR-CI are shown in Table \ref{tab2}. 

\begin{table}[!htb]
\caption{The results for model RVAE-FCNR-CI. We included an extra evaluation metric, mean confidence interval accuracy (MCIA), to assess the performance of our proposed method in estimating clinically useful 95\% confidence intervals. MRVA: mean relative volume accuracy. STD: standard deviation. R: Pearson’s correlation coefficient. MCIA: mean confidence interval accuracy. SEN: sensitivity. SPE: specificity. ACC: accuracy. }
\centering
\label{tab2}
\begin{tabular}{@{}cccccc@{}}
\toprule
\multicolumn{3}{c}{\textbf{Volume}}                 & \multicolumn{3}{c}{\textbf{Splenomegaly}}           \\ \midrule
\bm{$MRVA\pm STD$} & \textbf{R} & \textbf{MCIA} & \textbf{SEN} & \textbf{SPE} & \textbf{ACC} \\
\multicolumn{6}{c}{\textbf{Single view}}                                                         \\
$86.75\%\pm10.35\%$    & 0.9310     & 83.33\%       & 78.57\%      & 96.74\%      & 92.50\%      \\
\multicolumn{6}{c}{\textbf{Dual views}}                                                          \\
$92.14\pm 6.81\%$      & 0.9890     & 84.17\%       & 85.71\%      & 98.91\%      & 95.83\%      \\ \bottomrule
\end{tabular}
\end{table}

Note that the input to the fully connected layers of the RVAE-FCNR-CI model was a sample from the latent distribution instead of the latent representation mean. When estimating spleen volume (without sampling), the model showed a similar MRVA compared to the normal RVAE-FCNR model. The model trained with single-view data achieved a MCIA of 83.33\%, while the model trained with dual-view data achieved a MCIA of 84.17

\begin{figure}[!htb] 
\centering 
\includegraphics[width=1\textwidth]{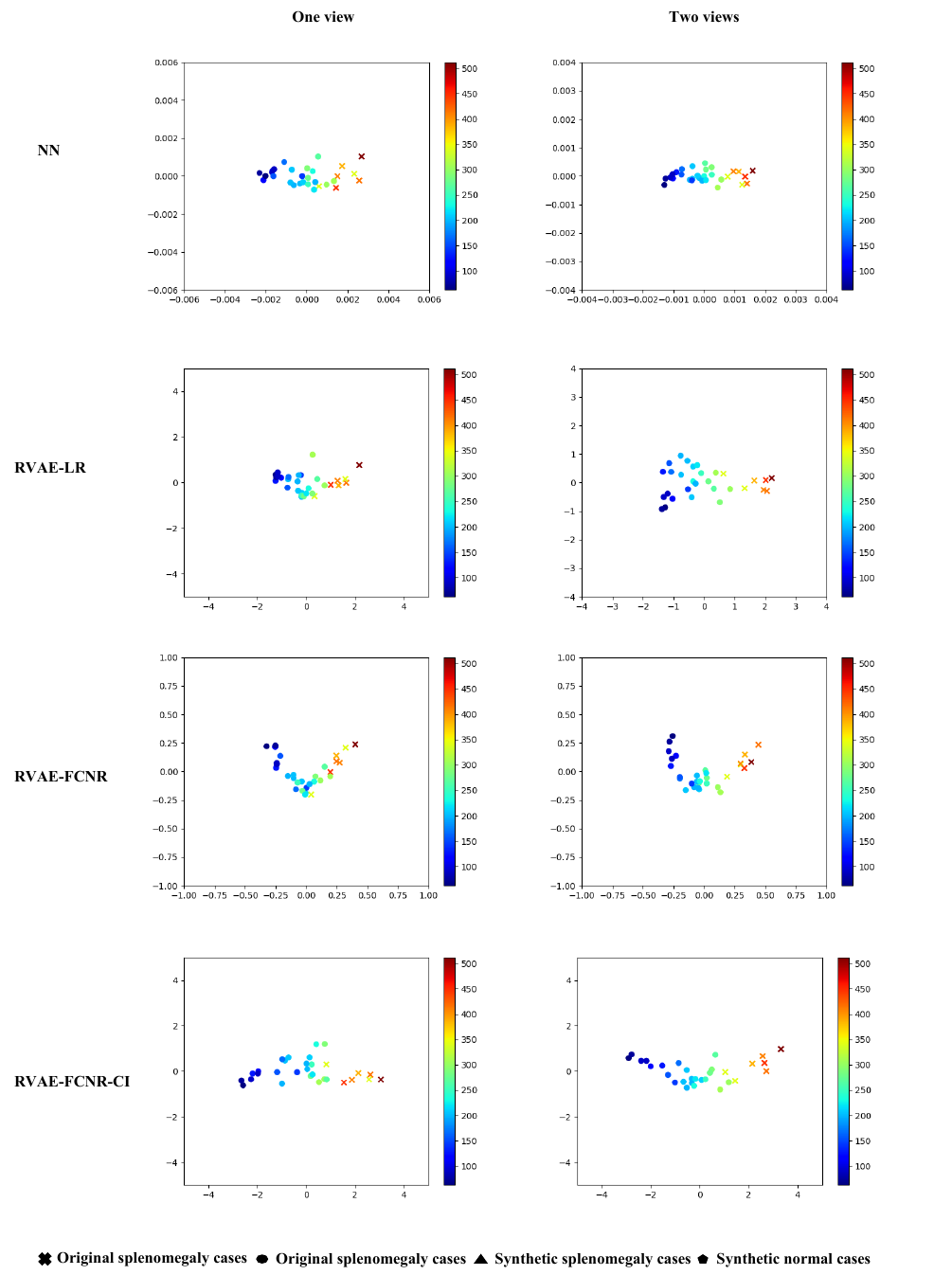} 
\caption{ Illustration of the test set latent spaces of our proposed models, i.e., NN, RVAE-LR, RVAE-FCNR and RVAE-FCNR-CI, trained with single-view data or dual-view data. The colours indicate estimated spleen volume according to the colour scales shown. The dots are normal spleen cases, and crosses are splenomegaly cases (i.e., with volume greater than 314.5mL).  } 
\label{fig7} 
\end{figure}

To visualise the latent spaces of all proposed models, we performed PCA on the latent representation mean and reduced its dimension from 128 to 2 for the test cases. The resulting 2-dimensional latent spaces are visualised in Fig. \ref{fig7}. 

\clearpage

It can be observed from the latent space plots of all our proposed models that normal cases and splenomegaly cases are clustered in different areas, with a clear boundary between them. In addition, there is a clear progression from “low volume” to “high volume” subjects, as indicated by the colour scale.  

\section{Discussion}
\label{section7}
In this work, we have proposed a novel framework utilising the power of VAEs to estimate spleen volume directly from 2D segmentations. Our overall objective is to use this framework to estimate spleen volume from clinically acquired 2D ultrasound data. The work presented here overcomes a major methodological challenge in this objective, i.e., estimating volume from segmented 2D slices. Combined with our previous work \cite{yuan2022deep}, this now opens up the possibility of fully automated spleen volume estimation from 2D ultrasound. In order to validate this overall pipeline, a dataset of 2D ultrasound with associated ground truth volumes would be required. Because of the limited availability of paired 2D ultrasound and 3D volumetric imaging datasets, in this paper we used 2D slices selected from CT spleen volume segmentations to develop our proposed models. In the future we plan to acquire a dataset that will enable us to evaluate the full pipeline, incorporating both automated 2D ultrasound segmentation and volume estimation. Ideally this should include high fidelity volumetric imaging such as CT as well as 2D ultrasound. 

We proposed three methods for obtaining spleen volume from the latent space distribution of the VAE. The NN method estimates the volume of unseen spleen images by finding the nearest neighbour in the latent space and using the corresponding volume as the estimation. The PLR method is based on a linear regression between embeddings in the latent space and spleen volume. Finally, the linear or non-linear relationship between the latent distribution embeddings and spleen volume is learned by RVAE-LR and RVAE-FCNR. As shown in Table \ref{tab1}, the accuracy of spleen volume prediction of the RVAE models is improved due to the increasing complexity of the model compared to NN and PLR. Our best method (RVAE-FCNR) outperforms human experts in estimating spleen volume by a significant margin (MRVA 86.62\% vs. 68.54\% from single-view data and 92.58\% vs. 81.60\% from dual-view data). 

In this work we chose to use a VAE-based approach to estimating volume. It would also have been possible to train a deep learning model to directly estimate volume without using a decoder as employed by the VAE. However, such strong supervision would likely result in a more biased representation than that learnt by the VAE, which (in its basic form) learns the representation in an unsupervised manner. Unsupervised representation learning (or less-supervised representation learning in the case of RVAE-LR and RVAE-FCNR) has the potential benefit of learning a representation that could be useful for other tasks. For example, it might be possible to use the framework we have developed to predict the need for clinical intervention and/or the likelihood of further spleen enlargement. In addition, the decoder of the VAE allows the reconstruction of virtual spleens from different parts of the latent space distribution, which could be used for gaining new insights into the relationship between spleen morphology and patient prognosis. We will investigate these possibilities in future work. 

The use of deep learning-based tools to improve existing clinical workflows has attracted increasing attention and has been widely investigated. However, the lack of interpretability in deep learning models has been one of the major concerns when applied to real clinical scenarios. Despite their accuracy, deep learning models remain black boxes, making it difficult for clinicians to understand how the outcome is obtained. This lack of interpretability can lead to a lack of trust in clinical decision-making. To alleviate this concern we have provided visualisations of the model’s latent space so that clinicians can see how the representation learnt by the model is closely related to volume. We observed that there is a clear boundary between splenomegaly cases and normal cases, which should improve clinician trust in the models. In addition, to provide more information to support clinicians, we developed a technique to quantify confidence intervals for the volume estimates, rather than just providing a single scalar value. In this approach, the regression layers made predictions from multiple samples in the latent distribution. This process is analogous to the uncertainty that arises from obtaining a 2D slice from a 3D spleen volume, where the resulting 2D segmentation can vary due to motion or slight differences in orientation. It is worth noting that our experiments on the use of confidence intervals was conducted solely on the RVAE-FCNR model. This is because the linear regression layer used in RVAE-LR is not constrained by activation functions, which can lead to negative volume predictions. With more real data, we believe it will be possible to represent a wider range of possible spleen shapes in the latent space, overcoming this problem with RVAE-LR. 

As mentioned in the Introduction, it would be possible to estimate spleen volume from 3D imaging such as CT. However, 3D imaging is usually not deployed to measure spleen volume as it is more expensive and less accessible. Practically, estimating the spleen volume from 2D imaging is more clinically useful as it can be incorporated into routine clinical workflows. In this setting, reconstructing the 3D segmentation volume from 2D segmentation slices and then calculating the spleen volume is a natural choice. Such a method was proposed and demonstrated in cardiac imaging by \cite{stojanovski2022efficient}. However, our results showed that the Pix2Vox++ 2D-to-3D reconstruction method did not perform as well as our direct volume estimation approach. 

We chose to use coronal slices of the spleen for our single-view setting. When ultrasound images of the spleen are acquired in a clinical setting the images acquired are almost always in this view, due to the limited acoustic windows. Therefore, this represents a clinically realistic scenario. We also investigated a dual-view setting which included transverse slices, which are also obtained in a typical ultrasound examination. 

Despite the potentially high diagnostic value of spleen volume, its clinical meaning remains under-investigated because of the difficulties involved in its estimation. We believe that our work could enable such an investigation and eventually result in more precise diagnoses of splenomegaly and diseases that can be evaluated using spleen volume.

\section{Conclusions}
In this work, we have proposed a VAE-based framework to estimate volume automatically from 2D spleen segmentations. To the best of our knowledge, this represents the first such work in the spleen and has surpassed human expert level and an existing reconstruction-based method in spleen volume estimation. 

\section{Acknowledgements}
This work was supported by the Wellcome/EPSRC Centre for Medical Engineering [WT 203148/Z/16/Z]. The support provided by China Scholarship Council during PhD programme of Zhen Yuan in King’s College London is acknowledged.

\bibliographystyle{ieeetr}

\end{document}